\documentclass[12pt]{tcibook}
\usepackage{fancyhea}
\usepackage{work}
\usepackage{bm}       
\usepackage{graphicx}
\usepackage{multirow}
\usepackage{lineno}
\usepackage{threeparttable}
\usepackage[normalem]{ulem}
\usepackage{color}
\usepackage[colorlinks = true,
            linkcolor = blue,
            urlcolor  = blue,
            citecolor = blue,
            anchorcolor = blue]{hyperref}
\usepackage[numbers]{natbib}
\usepackage{amsmath,amssymb}
\usepackage{enumitem}

\def\beq{\begin{equation}}
\def\eeq#1{\label{#1}\end{equation}}
\def\eeqn{\end{equation}}

\newenvironment{Eqnarray}%
   {\arraycolsep 0.14em\begin{eqnarray}}{\end{eqnarray}}
\def\beqa{\begin{Eqnarray}}
\def\eeqa#1{\label{#1}\end{Eqnarray}}
\def\eeqan{\end{Eqnarray}}

\let\bar=\overbar

\def\lsim{\mathrel{\raise.3ex\hbox{$<$\kern-.75em\lower1ex\hbox{$\sim$}}}}
\def\gsim{\mathrel{\raise.3ex\hbox{$>$\kern-.75em\lower1ex\hbox{$\sim$}}}}

\def\del{\partial}
\def\Dslash{\not{\hbox{\kern-4pt $D$}}}
\def\dslash{\not{\hbox{\kern-2pt $\del$}}}
\def\pslash{\not{\hbox{\kern-2pt $p$}}}
\def\ETmiss{\not{\hbox{\kern-4pt $E$}}_T}

\def\Dlr{\mathrel{\raise1.5ex\hbox{$\leftrightarrow$\kern-1em\lower1.5ex\hbox{$D$}}}}

\def\MSB{{\bar{M \kern -2pt S}}}
\def\msb{{\bar{\scriptsize M \kern -1pt S}}}

\def\drb{{\bar{\scriptsize D \kern -1pt R}}}

\setcitestyle{square}

\usepackage{comment}

\def\authorlist#1#2{
    \vskip 0.4in
\begin{center}\begin{large} {\bf  #1 } \end{large}
    \vskip 0.2in
              #2
     \vskip 0.2in
   \end{center}
}

\begin{document}


\pagenumbering{roman}

\parindent=0pt
\parskip=8pt
\setlength{\evensidemargin}{0pt}
\setlength{\oddsidemargin}{0pt}
\setlength{\marginparsep}{0.0in}
\setlength{\marginparwidth}{0.0in}
\marginparpush=0pt


\pagenumbering{arabic}

\renewcommand{\chapname}{chap:intro_}
\renewcommand{\chapterdir}{.}
\renewcommand{\arraystretch}{1.25}
\addtolength{\arraycolsep}{-3pt}

\setcounter{chapter}{5} 

\chapter{Report of the Topical Group on Dark Energy and Cosmic Acceleration: Complementarity of Probes and New Facilities for Snowmass 2021}
\chaptermark{\footnotesize Dark Energy and Cosmic Acceleration: Complementarity of Probes and New Facilities}

\authorlist{Conveners: Brenna Flaugher, Vivian Miranda, David J.~Schlegel}
   {Adam J.~Anderson, Felipe Andrade-Oliveira, Eric J.~Baxter, Amy N.~Bender, Lindsey E.~Bleem, Chihway Chang, Clarence C.~Chang, Thomas Y.~Chen, Kyle S.~Dawson, Seth W.~Digel, Alex Drlica-Wagner, Simone Ferraro, Alyssa Garcia, Katrin Heitmann, Alex G.~Kim, Eric V.~Linder, Sayan Mandal, Rachel Mandelbaum, Phil Marshall, Joel Meyers, Laura Newburgh, Peter E.~Nugent, Antonella Palmese, M.~E.~S.~Pereira, Neelima Sehgal, Martin White, Yuanyuan Zhang
}

\section{Executive Summary}

The mechanism(s) driving the early- and late-time accelerated expansion of the Universe represent one of the most compelling mysteries in fundamental physics today. The path to understanding the causes of early- and late-time acceleration depends on fully leveraging ongoing surveys, developing and demonstrating new technologies, and constructing and operating new instruments. This report presents a multi-faceted vision for the cosmic survey program in the 2030s and beyond that derives from these considerations.
%
%
Cosmic surveys address a wide range of fundamental physics questions, and are thus a unique and powerful component of the HEP experimental portfolio.

Wide-field surveys in the optical/near-infrared have played a critical role in establishing the standard model of cosmology, $\Lambda$CDM. 
We strongly advocate for continuing this extremely successful program into the coming decade and beyond.

Regarding photometric imaging surveys, the HEP community sees three options for Rubin Observatory beyond LSST, each of which would require different investments with costs and benefits needing detailed study. These studies must be undertaken a few years into the LSST so that the range of opportunities and trade-offs between them can be informed by the then-current scientific findings and open questions in the field.

The next generation of spectroscopic surveys has the opportunity to map a significant fraction of the observable Universe in three dimensions, tracking the expansion of the Universe and providing constraints on dark energy throughout most of cosmic history. The spectroscopic roadmap starts with continued operation of DESI (i.e., DESI-II), followed by a new wide-field spectroscopic facility that leverages and complements LSST imaging. \cite{2022arXiv220903585S}.

Observations of the cosmic microwave background (CMB) have provided one of the most powerful probes of the origin, evolution, and contents of our Universe.
Continuation of a strong CMB program will transform our understanding of the early Universe through measurements of tensor modes, test the particle content to unprecedented precision and provide unique insights about gravity, dark energy, and new physics through cross-correlation with the wide-field galaxy surveys advocated in this report.
HEP investment in CMB-S4 is critical to enable a diverse fundamental physics program.
Following CMB-S4, higher-resolution observations of the CMB will open a new regime of microwave background cosmology. 

Advancement of emerging techniques for cosmology and the study of dark energy, and complementarity among methods, should also be a priority.  An array of concepts for mapping the Universe using radio or millimeter-wave spectroscopy have promise as unique probes of large-scale structure.  Third-generation gravitational wave observatories now being studied have potential for independently probing the expansion of the universe and dark energy, which should be characterized and optimized. Across surveys and methods, priority should be given to the potential sensitivity gains from  joint processing.

This report arrives at several recommendations: 

\noindent {\bf Near-term Facilities}
\begin{itemize}

\item {Given the pivotal role of CMB experiments in the landscape of particle physics and cosmology, and their phenomenal successes thus far, we advocate for advancing the CMB program through strong support of the near-term  construction and operation of CMB-S4, which will cross critical, well-motivated thresholds in the searches for inflationary gravitational waves and new particle species.  }
\item We advocate for the continued operations of DESI (DESI-II;~\cite{2022arXiv220903585S}) as an important part of the spectroscopic roadmap while a Stage V spectroscopic facility is designed and built.

\item We advocate for support of small- and medium-scale projects that enhance the science reach of studies of transients discovered by Rubin LSST and ``standard sirens'' detected by gravitational wave facilities. Data from these projects should be combined with infrastructure that enables cross-experiment coordination and data transfer for time-domain astronomical sources and a US-HEP multi-messenger program with dedicated target-of-opportunity allocations on US-HEP and partner facilities.

\end{itemize}

\noindent {\bf Longer-term Facilities}
\begin{itemize}

\item  Through the Snowmass2021 process, the HEP community has identified the pressing need for next-generation wide-field, massively multiplexed spectroscopic capabilities to complement LSST imaging. We strongly advocate for the establishment, support and start of construction of a Stage V spectroscopic facility in the coming decade.

\item  Recognizing the wealth of fundamental physics that could be probed if much higher resolution and lower noise could be efficiently achieved over a wide-area CMB survey, we strongly advocate for support of studies of a Stage V CMB facility to bring it to conceptual readiness for the next decade. 

\item New approaches such as millimeter and 21-cm line-intensity mapping (LIM) hold the promise of exceptional cosmological constraining power.
However, the technological readiness of these programs must be further demonstrated before the community is prepared to invest fully in a large-scale project using these technologies.
Thus, we recommend a coordinated R\&D program to advance the technical readiness of these projects.

\item We advocate for the continued operation of the Rubin Observatory after LSST. The Rubin Observatory will continue to be a groundbreaking facility in 2034 that can advance the state-of-the-art by targeting the sky with new innovative observation strategies and/or instruments.

\end{itemize}

\noindent {\bf Complementarity}
\begin{itemize}[nosep]
\item No single experiment can reveal the nature of dark energy. 
Such a breakthrough will require data from a network of experiments, small and large, probing the early- and late-time Universe in complementary ways.
At present, cross-survey analyses are challenging to initiate, organize, and fund.  We advocate for the creation of clear pathways to support cross-survey analyses as part of the core mission of the HEP Cosmic Frontier.

\item Multi-messenger measurements of gravitational wave events are an emerging complementary technique for probing cosmology through standard sirens.  Support for coordination with future large facilities (such as the European Einstein telescope) will enable maturation of this novel technique for measuring dark energy.

\item We advocate for the creation of multi-site data archive centers, where data from cosmological surveys is replicated 
for robustness and continuous availability. The centers will provide the long-term preservation of datasets and simulations. Such centers should also supply computing resources for in-place analyses, 
making joint investigations attainable given the huge I/O bottleneck that arises from downloading data from such centers.

\item We advocate for a robust program to increase the available supercomputing resources to enable running, postprocessing, and validating a diverse set of numerical gravity-only and hydrodynamical simulations tailored to the specificities of different surveys. This program would enable the running and testing of data-driven methods involving, for example, machine learning or bayesian methods.

\end{itemize}

\begin{figure*}[t]
\includegraphics[width=6.5in]{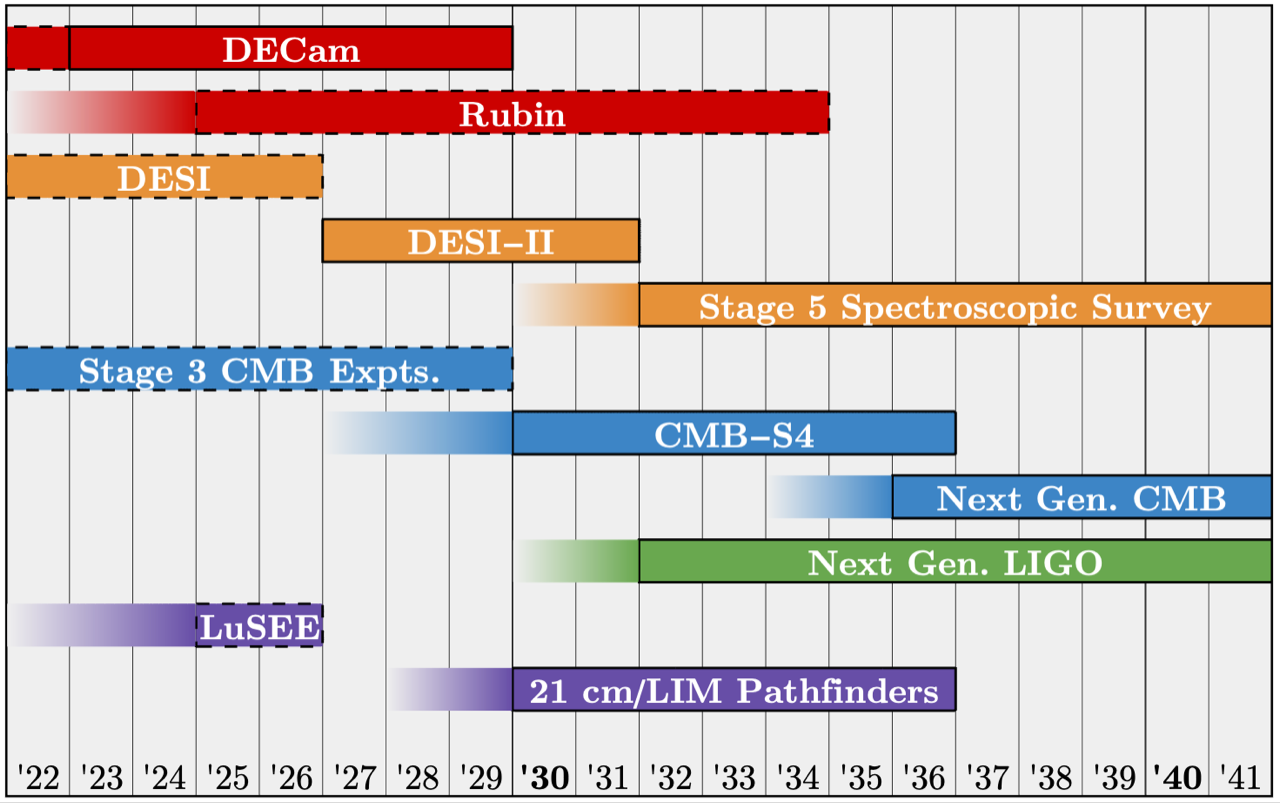}
\caption{Current and potential future facilities probing cosmic acceleration that are or may be supported by DOE or NSF. Dashed boxes indicate fully-funded facilities. Facilities in red are optical imaging, in orange are optical spectroscopy, in blue are CMB, in green are gravitational waves,  and in purple are radio/mm spectroscopy.  The fade-in regions indicate commissioning periods, while the boxes indicate full survey observations.} 
\label{fig:roadmap}
\end{figure*}

\section{Introduction}

Cosmic surveys, including observations of the cosmic microwave background (CMB) and the distribution of stars and galaxies, enable investigations of the fundamental components of the Universe including dark energy, dark matter, inflation, the properties of neutrinos, and signatures of other ``dark sector'' particles.
Cosmological and astrophysical measurements provide the only empirical measurements of dark energy and inflation, while measurements of dark matter and neutrinos both motivate and complement other terrestrial HEP experiments. 
Over the last several decades cosmic surveys have resulted in the creation of a ``Standard Model" of cosmology ($\Lambda$CDM), in which the Universe is currently comprised of ${\sim} 68\%$ dark energy (assumed to be a cosmological constant, $\Lambda$) and ${\sim}27\%$ non-baryonic, collisionless, cold dark matter (CDM)~\citep[e.g.,][]{SDSS:2005xqv,DES:2018,DES:2019,eBOSS:2020yzd,DES:2022,SPT-3G:2021eoc}. The fact that cosmic surveys can address a wide range of fundamental physics questions make them a unique and powerful component of the HEP experimental portfolio.
\begin{figure*}[ht]
\includegraphics[width=6.5in]{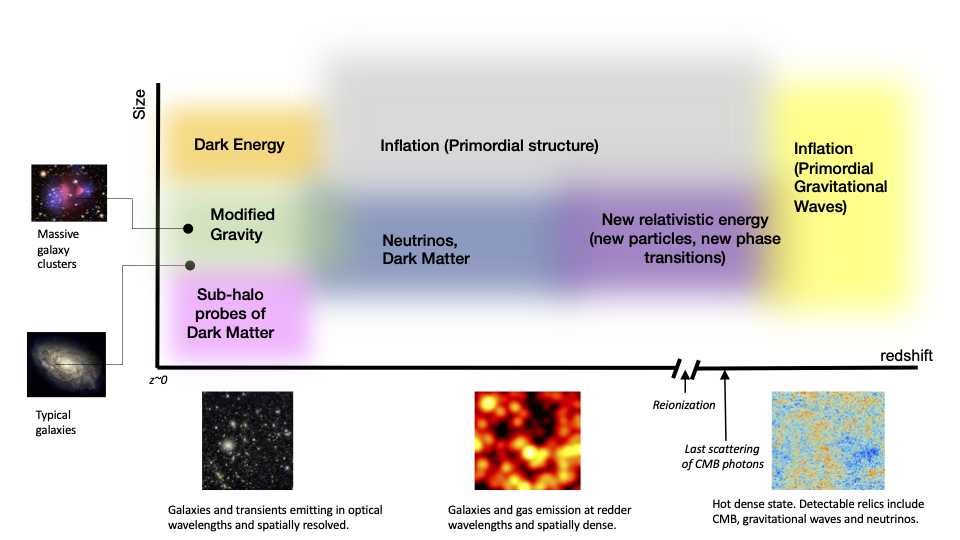}
\caption{Key scientific opportunities in HEP targeted by cosmic survey facilities. The colored areas illustrate regions in spatial scale and redshift favored for various scientific targets. Dark energy and modified gravity favor measurements at lower redshift at large-to-moderate spatial scales.
Inflationary signals are best explored at the largest spatial scales.
Small-scale, low-redshift surveys explore dark matter in the sub-halo regime, and precision measurements of the matter power spectrum at moderate-to-small scales out to high redshift are sensitive to neutrinos, dark matter, and new relativistic energy in the early Universe. 
Each independent technique explores physics over a broad range of spatial scales and cosmic history, while the full suite has multiple complementary measurements providing robust results.
} 
\label{fig:spatial_redshift}
\end{figure*}

Figure \ref{fig:spatial_redshift} shows the breadth of HEP scientific opportunities enabled by cosmic surveys, stretching from the earliest moments of the Universe to present day.    The previous P5 science driver of ``Understanding dark energy and cosmic acceleration'' is still very relevant and will continue to be so for the next decade and beyond.   The theorized epoch of inflation is shown at the highest redshifts and optimally probed through tracers at the largest spatial scales on the sky.  In contrast, dark energy is shown at the left, as its impact is most significant on the growth of structure in the modern (late-time) universe.  Dark energy is optimally probed at large to medium scales.   Cosmic signals that probe inflation and dark energy  include (but are not limited to) the cosmic microwave background from the very early universe, the gas and galaxies tracing the matter distribution as structure formed and evolved, and optical galaxies and transients in the late universe.  Each of theses signals has unique strengths that are discussed in further detail below.  Additionally, cross-correlating between cosmic signals can eliminate systematics and extend the scientific reach further than that of the individual measurements.  Finally, it is important to note the wealth of physics beyond dark energy and inflation that these very same cosmic signals can probe.   From neutrinos and new relativisitic particles, to modified gravity and dark matter, cosmic signals have the ability to answer some of the biggest questions currently facing high-energy physics. 


\begin{figure*}[t]
\begin{center}
\includegraphics[width=4.5in]{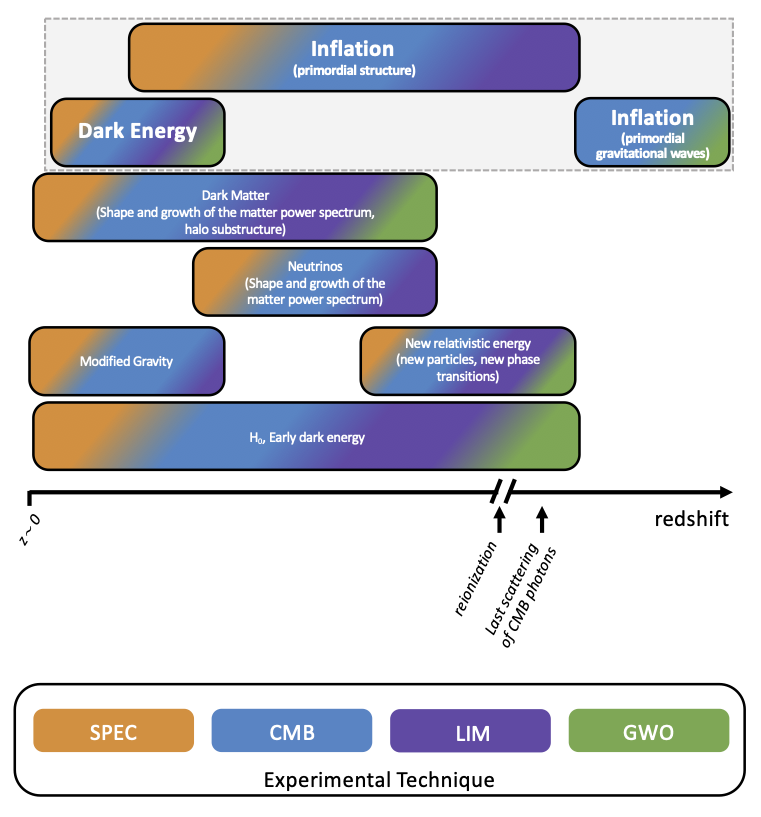}
\caption{A high-level summary of the key scientific opportunities.  The horizontal extent of each box corresponds to the redshift-range of the tracer, while the coloring indicates the experimental technique used to measure the signal. The dashed grey box emphasizes dark energy and inflationary probes. 
\label{fig:redshift_complement}}
\end{center}
\end{figure*}

Figure \ref{fig:redshift_complement} shows a simplified summary of these same scientific targets as well as the cosmic signals and techniques used to explore them.  Four main techniques are shown and discussed further in this report. First, optical and near-infrared surveys combining both imaging and spectroscopic (SPEC) measurements to measure tracers of structure (such as galaxies) in the late-time universe.  Section \ref{sec:optical} discusses the highly anticipated scientific impact of the Vera Rubin Observatory (LSST) and the Dark Energy Spectroscopic Instrument (DESI), as well as complementary facilities and a future envisioned Stage V spectroscopic survey.  Next,  Section \ref{sec:cmb} introduces CMB facilities, including the CMB-S4 experiment that was prioritized in the previous Snowmass and P5 process.  Also discussed is one concept for a CMB facility that is a potential successor to CMB-S4.   Section \ref{sec:cross} highlights the power of cross-correlations between optical imaging, spectroscopic, and CMB surveys.   This section also introduces transients and gravitational wave observations (GWO, emitted from local universe sources)  as probes of fundamental physics, which also relies on complementary survey observations.    Smaller projects and technology pathfinders are described in Section \ref{sec:smallproj}.  Included is a discussion of line-intensity mapping (LIM) both using the 21-cm line from neutral hydrogen and mm-wavelength tracers, such as the rotational transitions of CO and the [CII] ionized carbon fine structure line.  Finally, Section \ref{sec:multimessenger} details current and future gravitational wave observatories that will provide gravitational wave events for multi-messenger probes.    Altogether, these observational techniques and cosmic survey facilities provide a unique and powerful means to explore dark energy and inflation in the coming decade, as well as developing the technology and concepts needed to continue a vibrant and cutting-edge program in the years that follow. 

\section{Optical/Near-Infrared Surveys and Facilities}
\label{sec:optical}

\begin{figure*}[t]
\centering\includegraphics[width=6in]{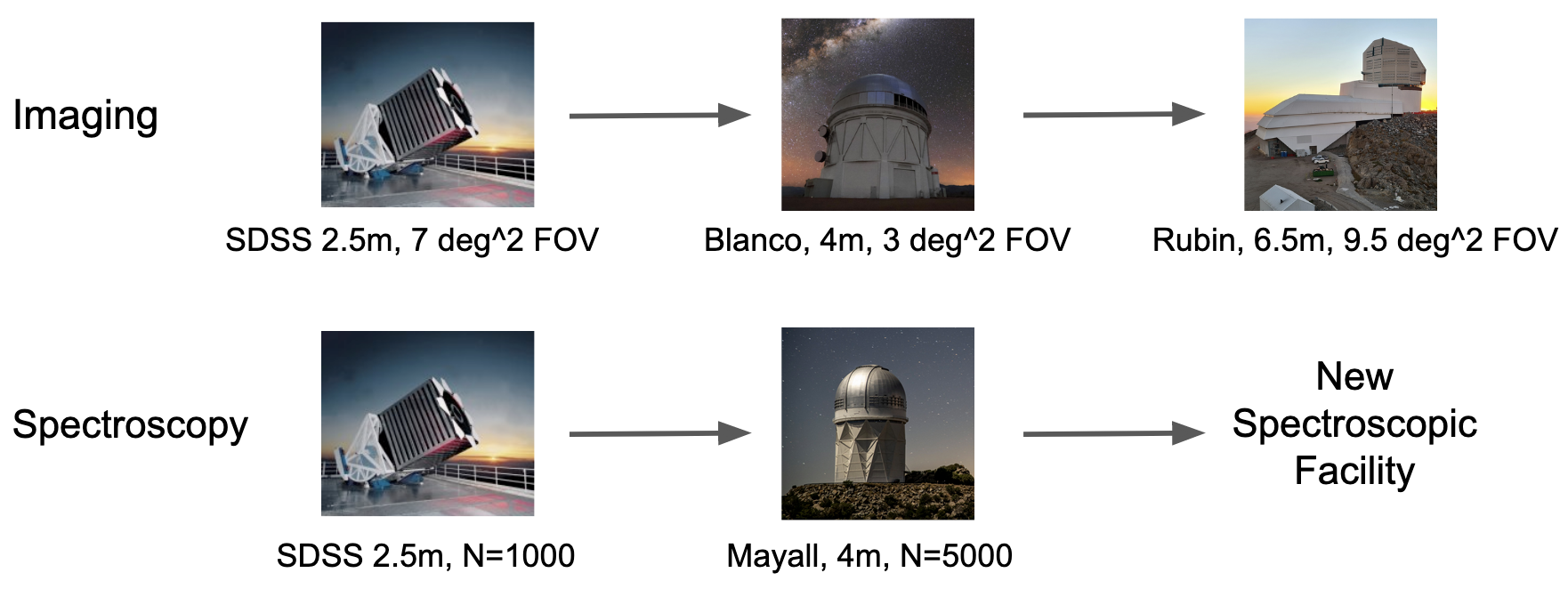}
\caption{Summary of imaging and spectroscopic surveys and facilities, ongoing and planned, that are supported by DOE/NSF partnerships. The international ground and space-based landscape of optical wide-field surveys, ongoing and planned, is very rich but for clarity is not represented here. SDSS had both imaging and spectroscopic capabilities, the Blanco telescope was used to carry out the DES, and the Mayall is currently used for DESI. In the near future, the Rubin Observatory will begin LSST. A new spectroscopic facility would open up new scientific opportunities.} 
\label{fig:opt_facilities}
\end{figure*}

Wide-field surveys at optical and near-infrared wavelengths play a central role in the exploration of the physics of the dark Universe. The Sloan Digital Sky Survey (SDSS), the first major survey jointly supported by the DOE and NSF, delivered unprecedented measurements of the structure of the Universe at late times. SDSS had first light in 1998 and provided both imaging and spectroscopic data. DOE-supported upgrades to the instrumentation in 2007--2009 enabled the cosmology reach to earlier cosmic times with the SDSS-III/BOSS and SDSS-IV/eBOSS programs. BOSS and eBOSS were spectroscopic surveys focused on refining measurements of the BAO signal through extensions of the SDSS program. 

Building upon the tremendous success of SDSS, new optical surveys have been designed, constructed, and executed through continued partnership between DOE and NSF. The Dark Energy Survey (DES) is an imaging survey that was operated on the 4-m Blanco telescope in 2013--2019 and is currently extracting final cosmology results.  DES has delivered exciting results on the fundamental physics of dark energy, modified gravity, and dark matter. The Rubin Observatory is under construction in Chile and will start the Legacy Survey of Space and Time (LSST) in 2024. LSST will survey  the southern sky with an unprecedented combination of depth, visit frequency, spectral bands, and areal coverage to provide unprecedented constraints on dark energy, neutrinos, and dark matter over the course of its 10-year survey. Recently, the Dark Energy Spectroscopic Instrument (DESI) started its observational campaign on the 4-m Mayall telescope in pursuit of measurements of dark energy, neutrino mass, and dark matter. 

Wide-field surveys in the optical/near-infrared have played a critical role in establishing the standard model of cosmology, $\Lambda$CDM and have delivered a broad range of science in addition to dark energy studies. 
This exceptional success showcases the power of imaging and spectroscopic surveys, and we strongly advocate for continuing this extremely successful program into the coming decade and beyond. 
In particular, the unparalleled efficiency of DESI for wide-field spectroscopy and the unprecedented imaging survey data to be collected by the Rubin Observatory will open up many exciting directions for advances in cosmology. 

In the following, we provide first a brief summary of facilities that are currently operating (DESI) or will soon start operations (Rubin Observatory). Then we discuss future opportunities with either existing or new facilities. We emphasize the following priorities for the optical survey program:

\begin{itemize}
\item Support for extracting science from ongoing and near-future surveys;
\item Support for small programs that use existing facilities to maximize the science from flagship facilities;
\item Support for the development of new technology to enable future surveys;
\item Support for the design and development of a Stage V spectroscopic survey.
\end{itemize}

\subsection{Rubin Observatory}

The Vera C.\ Rubin Observatory is a powerful facility that will further our knowledge of the Universe in many ways by enabling studies of the nature of dark energy and dark matter, a deep census of the solar system, exploration of the transient optical sky, and surveys of the stellar populations of the Milky Way
~\cite{2019ApJ...873..111I}. The Legacy Survey of Space and Time (LSST) to be undertaken with the observatory is due to start operations in 2024 and map the Southern sky for 10 years. LSST will deliver exciting science opportunities and we stress that support for LSST science will be crucial for the community.
Precursor surveys have shown that data from a new survey always come with unexpected challenges but also opportunities. To address the challenges and to take advantage of new opportunities, sufficient support of the science programs is essential.

After LSST is completed, Rubin will still be a state-of-the-art survey facility. The
Rubin White Paper~\cite{Blum:2022dxi} describes possibilities for future endeavors for the observatory, and provides the scientific motivations for three post-LSST scenarios. Given that this CF6 report focuses on future facilities, we summarize them here and refer the reader to the White Paper and the CF4 report for the scientific justifications.   

The post-LSST opportunities for Rubin are in three broad categories, as described in Ref.~\cite{Blum:2022dxi}:
\begin{itemize}
    \item \textbf{Continuing operations}: A strong science case for continued cooperation of Rubin relates to time-domain studies that would rely on modified observing cadence, exposure time, or filter selections relative to the LSST survey for greatly enhanced efficiency and target-of-opportunity observations of rare phenomena. Other scientific cases for continued operation of the observatory relate to follow-up observations of discoveries with LSST, focusing on studies that would enhance understanding the fundamental nature of dark matter. Continuing operations of Rubin,  modifying only the observing strategy, could also provide synergistic observations that enable better scientific outcomes from combined analyses with overlapping large-area deep optical surveys in support of  cosmology. In particular, the planned 2000 deg$^2$ High Latitude Survey with the Nancy Grace Roman Observatory would be an important target. 
    \item \textbf{New filters}:  Several scientific opportunities would be enabled by installation of new photometric filters. Examples discussed in Ref.~\cite{Blum:2022dxi} include a filter set complementary to the original six to improve photometric redshift estimates of the catalogued galaxy sample; a set of narrow-band or medium-band filters to enable emission line surveys for particular lines at redshift $z = 0$ or to select samples of galaxies at a set of discrete redshifts; and a set of patterned filters, which would enable multiple bandpasses to be sampled simultaneously across the field. 
    \item \textbf{New instrument}: This would be the most expensive option but could transform the Rubin Observatory by providing truly new capabilities. For example, a wide-field spectrograph would provide the opportunity to follow up the rich LSST imaging dataset and open many new scientific approaches. This option would require a detailed feasibility and design study in the near future.
\end{itemize}

\subsection{Dark Energy Spectroscopic Instrument} 

The next decade promises exciting findings to gain a better understanding of the physics of the dark Universe. DESI, located on the 4-m Mayall Telescope at Kitt Peak, Arizona~\cite{desi:2016,desi2:2016}, is the first Stage IV dark energy experiment to begin science operations. 
DESI consists of a focal plane with 5,000 fiber positioners, a field-of-view with a diameter of 3.2 deg and ten 3-channel spectrographs covering the wavelength range 0.36--0.98 $\mu$m.  
DESI is currently conducting a 5-year survey to measure redshifts of 40 million galaxies plus a survey of gas in the intergalactic medium to constrain dark energy and cosmological parameters using the BAO and RSD techniques. At the end of the survey in 2026, the instrument will still be competitive with all other multi-object spectrographs that will exist at the time.  The proposed DESI-II survey would continue operating the instrument (possibly with upgrades) leveraging and complementing the first
year or two of imaging data from Rubin LSST.
Additional spectroscopic data can enhance Rubin science in several ways (e.g., in photometric redshift training).  
Additionally, the DESI instrument is being considered as a possible contributor to Snowmass CF4 programs, particularly a large-volume survey to study inflation, neutrinos, and early dark energy in the linear/quasi-linear regime~\citep{Ferraro:2022}, and a large number density survey to study dark matter physics, modified gravity, small scale features in the primordial power spectrum, and possibly unknown physics~\citep{Dawson:2022}.

The continued operation of the DESI instrument (DESI-II) is an important first step in the future spectroscopic roadmap and it is currently at the early stages of conceptual design \citep{DESI2:2021}.
Several unique science opportunities are possible, either by continued operations of the current instrument, or with modest technological upgrades. These include dense surveys of the local volume for precision measurements of dark matter, dark energy, and high-resolution studies of the cosmic web (and transients followup for gravitational waves, supernovae, etc.), extension of the Luminous Red Galaxy (LRG) and  Emission Line Galaxy (ELG) samples to higher redshift to enable multi-tracer analyses and take advantage of sample-variance cancellation, as well as increasing the observed volume, allowing access to larger scales, providing the cleanest probes of primordial physics.

A high-redshift ($z > 2$) survey of Lyman-alpha emitters (LAEs) and Lyman Break Galaxies (LBGs) would measure a volume comparable to the main DESI samples, but at a different cosmic time. This would allow measurements of the amplitude of fluctuations at high redshift, a particularly compelling measurement in light of the recent tensions between the amplitude of structures at late time ($z<1$), compared to the predictions from the CMB (the so-called ``$S_8$ tension''). Measurements in the intermediate redshift regime ($2 \lesssim z  \lesssim 4$) are particularly well-suited for understanding the origin of this tension. Moreover, measurements of expansion over this redshift range, deep into the matter-dominated epoch, will shed light on dynamical dark energy, where many models mimic a cosmological constant at late times, but can differ significantly from it during matter domination.
In addition to its science reach, such a survey would also serve as a pathfinder for extended wide-field observations of high-redshift galaxies by a future facility, as discussed in the next Section.

Possible technology upgrades to the DESI instrument include replacement of detectors with low-read-noise Skipper CCDs, a replacement of the focal plane with a larger number of fiber positioners, and the addition of a 4th spectroscopic channel extending further into the IR to measure [OII]-emitting galaxies to $1.6<z<2.0$, not currently possible with the existing 3 channels.

Moreover, the potential overlap between DESI and LSST is an impressive 14,000 square degrees of extragalactic sky if both instruments were to observe to their design limits ($-30^{\circ} < {\rm Dec.} < +30^{\circ}$) and represents a great opportunity to complement LSST observations with galaxy spectroscopy. The most ambitious upgrade of DESI would include the replacement of the primary mirror, effectively turning the instrument into MegaMapper, a candidate future stage V spectroscopic facility described in the next Section.

\subsection{Stage V Wide-field Multi-Object Spectroscopy} 

By 2030, Rubin LSST will have mapped at least $\sim$20,000 deg$^2$ of the sky at unprecedented depth from Cerro Panch\'{o}n in Chile. 
LSST will measure the expansion history and structure of the Universe through observations of type Ia supernova, weak lensing, galaxy clustering, strong lensing, and ultra-faint galaxies.
However, LSST provides only coarse spectral information, and spectroscopic capabilities are essential to maximize the fundamental physical output from cosmic surveys~\citep{Kavli:2016}. 
Current wide-field spectroscopic capabilities in the southern hemisphere are insufficient for the task of complementing Rubin LSST. 
Existing capability is dominated by the Anglo-Australian Observatory's 2dF, with 400 optical fibers covering $\sim$3 deg$^2$ field-of-view on the 3.9-m AAT in Australia. The 4MOST instrument~\citep{4MOSTSPIE}, currently under construction and scheduled to begin operations soon, will measure $\sim$2400 spectra simultaneously using the 4-m VISTA telescope at the European Southern Observatory. Larger instruments, such as the 6.5-m Magellan telescopes at Las Campanas Observatory, the 8-m Gemini Telescope at Cerro Pach\'{o}n, and the 8.2-m Very Large Telescope at the European Southern Observatory (all in Chile), have fields-of-view that are too small for wide-field surveys.  Other facilities are planned with 8-m to 30-m mirrors, but also have fields-of-view that are insufficient for large-field surveys. 

The Snowmass2021 Cosmic Frontier is charged with synthesizing community input on future studies of dark energy, dark matter, inflation, neutrinos, and other light relics through observational cosmology  within the HEP program. 
Through the Snowmass2021 process, the HEP community has identified the pressing need for additional wide-field spectroscopic capabilities to complement LSST imaging~\citep{Dawson:2022,Ferraro:2022}. 
Understanding of the needs has evolved from previous community studies on maximizing science from LSST in 2015--2016~\citep{NAS:2015,Kavli:2016} and from the HEP Cosmic Visions process in 2016--2018~\citep{Dodelson:2016a,Dodelson:2016b,Dawson:2018fob}.
Several white papers have been submitted to Astro2020 and Snowmass2021 describing the physics program and facilities that could meet some or all of these needs including DESI-II~\citep{DESI2:2021}, MegaMapper~\cite{Schlegel:2019eqc,Schlegel:2021,2022arXiv220904322S}, the Maunakea Spectroscopic Explorer (MSE)~\cite{Marshall:2019wsa,Marshall:2021}, and SpecTel \citep{Ellis:2019gnt}. In-depth discussion of the science opportunities, together with detailed forecasts for a number of experimental configurations have been presented~\cite{Ferraro:2022, Sailer:2021, MegaMapper_science:2020}.

The fundamental physics program of a future spectroscopic facility is diverse and multifaceted. Following the  evolutionary history of the Universe from early to late times:

\begin{itemize}
\item {\bf Inflation}: A next-generation spectroscopic survey will access an extremely large volume of the Universe, which will enable it to measure a number of primordial quantities beyond the cosmic variance limit of the CMB. These include making exquisite measurements of the power spectrum, dramatically increasing the sensitivity to primordial features or oscillations that can be created by many models of inflation.
Sharp features arise when there is a sudden transition during inflation such as a step in the potential. Resonant features arise when some component of the background oscillates with a frequency larger than the Hubble scale. 
Another important advance achievable by these surveys is measurement of primordial non-Gaussianity with the goal of an order-of-magnitude improvement in sensitivity to surpass $\sigma(f_{\rm NL}^{\rm local}) < 1$, allowing the two main inflationary scenarios (single field vs multi-field inflation) to be distinguished. 
Additionally, greatly improved measurements of the running of the spectral index and of spatial curvature will shed additional light on the physics of the early Universe. 

\item {\bf Neutrinos and Dark Radiation}: Measurements of the physics of the early Universe provide strong constraints on the dark sector via, for example, via the determination of the number of light particles that are thermalized. This is parameterized by $N_{\mathrm{eff}}$, the number of relativistic particles other than photons. The Standard Model with three neutrino species predicts $N_{\mathrm{eff}}=3.045$. 
Measurements of the matter power spectrum can detect or exclude the existence of other particle species that decouple after the QCD phase transition, and tightly constrain particles that decouple earlier.
Cosmological measurements from large galaxy surveys will complement CMB observations and other experimental efforts to detect low-mass dark sector particles (e.g., via quantum sensors, a 3\,GeV muon beam dump experiment, and DarkQuest).

\item {\bf Dark Energy Throughout Cosmic History}: We are now in the domain of precision tests of the $\Lambda$CDM model. 

During this decade, experiments like DESI, Rubin LSST, Euclid, and the Roman Space Telescope will map the expansion of the Universe up to redshifts of $z \sim 2$ (when the Universe was roughly one-third of its current size).  A wide-field multi-object spectroscopic facility is needed to map the expansion of the Universe to higher redshifts (earlier times). A detailed 3D map of at least $\sim 40$ million galaxy positions with redshifts in the range $2 < z < 5$ is needed to take the next step in dark energy research. Precision measurements of the redshifts of $>40$ million distant galaxies will require an increase of about an order of magnitude in the combination of the number of fibers and light collection capabilities over current spectroscopic instruments, driving the design of future facilities. Additionally,  precision measurements of the matter power spectrum will be able to provide indirect percent-level constraints on Early Dark Energy (EDE) up to $z \sim 10^5$, when the Universe was only a few years old~\cite{Sailer:2021}.
 
\end{itemize}

\subsection{Complementary Facilities}

The optical/near-infrared dark energy facilities described in this section will be complemented by several ground- and space-based observatories at similar wavelengths.
They will be in various phases of planning, construction, and operation over the coming decades.
Since these facilities are currently driven by support from NASA, NSF-AST, and private contributions, we summarize them briefly here. 
We note that future support from DOE or NSF-PHYS could come through future instruments, US Extremely Large Telescopes (US-ELTs) or support for joint analyses. 

\begin{itemize}
\item {\bf US Extremely Large Telescopes} The US-ELT program consists of two 30-m-class telescopes: the Giant Magellan Telescope (GMT) to be sited in Chile and the Thirty Meter Telescope (TMT) to be sited in Hawai`i.
These telescopes have relatively small fields-of-view and multiplexing, and thus are not optimal as wide-area  spectroscopic survey facilities.
However, the large light collecting area provided by a 30-m mirror allows these telescopes to observe extremely faint objects quickly.
The US-ELT program was the highest-ranked ground-based porgram in the Astro2020 Decadal survey, but it is unlikely that the HEP community will participate in the design or construction of these telescope facilities.
However, 
US-ELTs could complement one of the surveys discussed in this section by providing, for example, deep spectroscopy for training photometric redshift estimators on the faintest galaxies observed by Rubin or high-resolution imaging data to constrain dark matter through strong lensing.
The cost of an ELT instrument (${\sim}\$40$M) would be roughly comparable to the cost of other HEP cosmic survey construction projects (e.g., DECam or DESI).

\item {\bf Small, Wide-field Optical Surveys} Both the Zwicky Transient Facility (ZTF) and the La Silla Schmidt Southern Survey (LS4) provide a complementary, and necessary, set of observations to those of the Rubin and the space-based surveys. ZTF and LS4 have direct relevance to several cosmology and fundamental physics efforts including: peculiar velocity measurements, and hence fundamental constraints on general relativity, with supernova as standardized candles; gravitational wave standard sirens as probes of the expansion of the
Universe and gravity; and measurements of the Hubble constant through Type Ia and II-P supernovae. They provide a higher cadence than the aforementioned surveys, especially important for analyzing the light curves as well as triggering follow-up for low-$z$ supernovae, and both have a robust ToO program for GW counterpart discovery in the optical. In addition, they open up the possibility of improved calibration for both Tully-Fisher and Fundamental Plane measurements (from spectroscopic surveys such as DESI) via supernova distances.

\item {\bf Space-based Observatories} \textit{Some description of Euclic, Roman, SpherEx and any others...}

\item {\bf Gravitational Wave Observatories} \textit{Some description of LIGO, Cosmic Explorer,...}

\end{itemize}

\section{Cosmic Microwave Background Surveys}
\label{sec:cmb}
Wide-field surveys of the CMB play a central role in particle physics and cosmology.  Missions such as {\it{COBE}} \cite{Mather:1982}, {\it{WMAP}}~\cite{Bennett:2013}, and {\it{Planck}}~\cite{Planck:2011} have provided critical insight into the birth and early evolution of our Universe.  In addition, many ground-based CMB experiments such as AdvACT~\cite{advact:2016}, SPT-3G~\cite{sobrin:2022}, BICEP/KECK~\cite{BK:2022}, and Simons Array~\cite{SA:2016} continue to push the frontiers of CMB measurements into lower-noise and higher-resolution regimes.  Power spectra of CMB temperature and polarization data provide some of the tightest constraints on particle physics models, dark matter, and inflation, and -- combined with measurements of gravitational lensing of the CMB -- compelling evidence for dark energy. 

\begin{figure*}[t]
\includegraphics[width=6.5in]{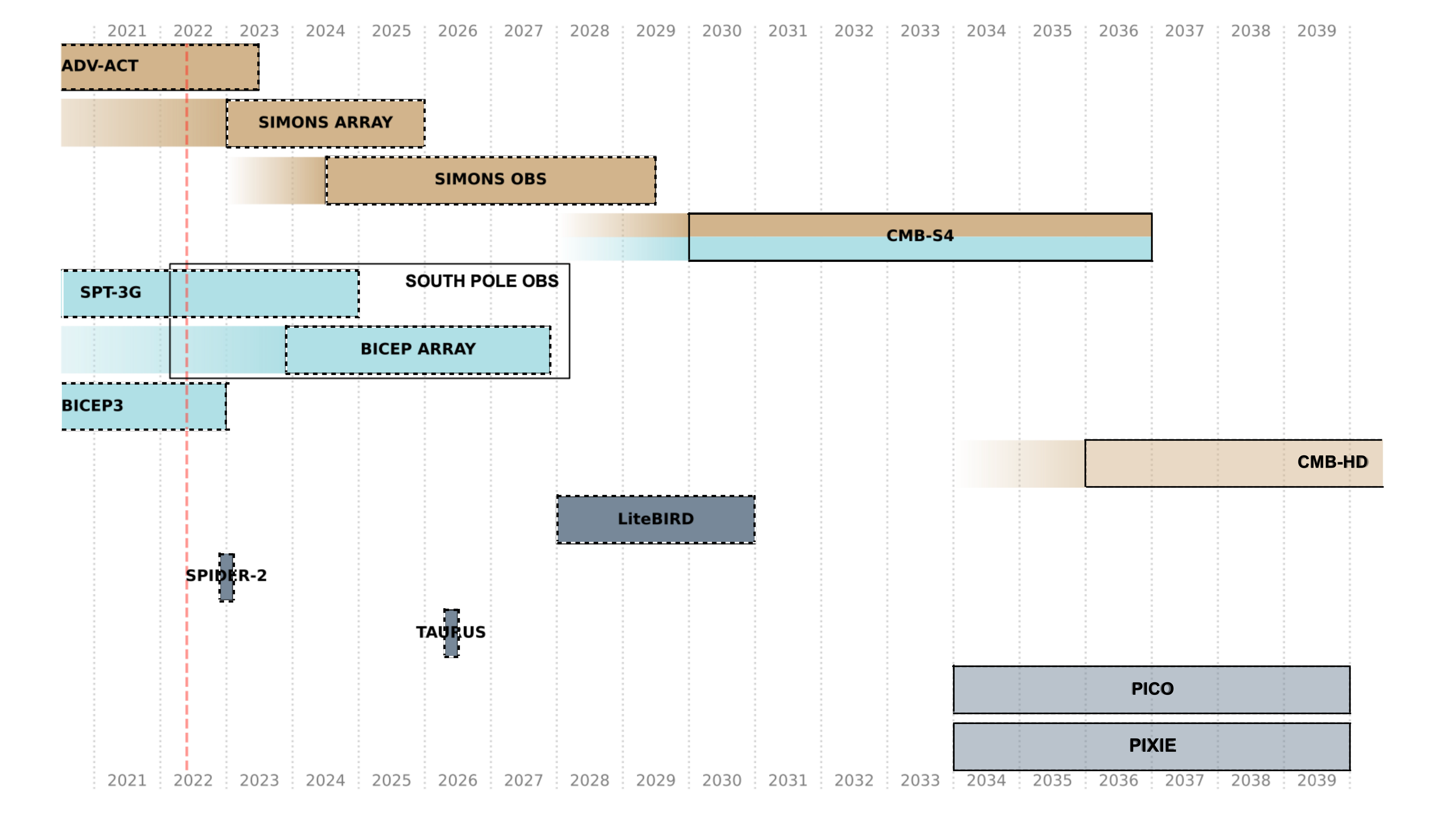}
\caption{Timeline of current and future ground-based CMB experiments. For context, the timeline also includes a few sub-orbital and satellite experiments in grey. Dashed boxes indicate fully-funded facilities.  The fade-in regions indicate commissioning periods, while the boxes indicate full survey observations.} 
\label{fig:CMB-timeline}
\end{figure*}

Building on the successes of precursor CMB experiments, the Simons Observatory~\cite{Simons:2019} and the South Pole Observatory are commencing observations in the early 2020s (see Figure~\ref{fig:CMB-timeline}).  Looking ahead, the CMB-S4 project~\cite{CMB-S4:2022ght} is planned to make significant leaps in sensitivity.  CMB-S4 is a joint DOE and NSF project that received DOE CD-0 approval in 2019, is advancing toward DOE CD-1 and NSF PDR, and currently has broad engagement from the majority of the US ground-based CMB science community. On a longer time-scale,  CMB-HD is a proposed experimental concept that would have six times the resolution of current and planned CMB experiments, opening up a new regime of millimeter-wave science~\cite{CMB-HD:2022bsz}.

Given the pivotal role that CMB experiments play in the landscape of particle physics and cosmology, and their phenomenal successes thus far, we strongly advocate for continuing the CMB program into the coming decade and beyond.  Similar to the optical survey program, we emphasize three priorities for the CMB survey program:

\begin{itemize}
\item Support for ongoing and near-future surveys, including CMB-S4;
\item Support for the development of new technology to enable the next major survey post CMB-S4;
\item Support for the design and development of the next major survey.
\end{itemize}

\subsection{CMB-S4} 

CMB-S4 is the  next-generation (Stage IV)  cosmic microwave background experiment~\cite{CMB-S4:2022ght}.  CMB-S4 is designed to achieve an enormous increase in sensitivity compared to existing CMB experiments while simultaneously leveraging two premier observing sites.  Combined, these unique features will enable CMB-S4 to make transformational measurements of primordial gravitational waves and inflation and the dark Universe~\cite{cmbs4:2019}.  Both of these science themes are of significant interest to the high-energy physics and cosmology communities.   Additionally, the unique properties of CMB-S4 will enable mapping the matter in the cosmos and studies of the time-variable millimeter wavelength sky.  CMB-S4 is a joint DOE and NSF project that has strong community support as evidenced by the mature, large collaboration and endorsements in the previous Snowmass, P5 report~\cite{HEPAPSubcommittee:2014bsm}, and more recently the Astro2020 Decadal Survey Report~\cite{Astro:2020}.

CMB-S4 will construct telescopes in both Chile and at the South Pole, taking best advantage of features of each site to pursue its scientific goals.  The South Pole site will host 18 small-aperture telescopes (SATs, diameter ~ 0.5 meter) and one 5-meter large-aperture telescope (LAT).  These telescopes will conduct an ultra-deep survey of 3\% of the sky, targeting the B-mode polarization measurements at both the large and small angular scales on the sky needed to constrain inflation.    Two 6-meter LATs will conduct a deep and wide survey of 60\%  of the sky from the Chilean site, targeting the CMB-S4 science goals that benefit from additional sky area.  Over 550,000 detectors will be deployed across the CMB-S4 telescopes, an enormous increase over all Stage III experiments combined that will make the planned increase in sensitivity possible.  As introduced above, CMB-S4 has four main science themes that drive this experiment design and subsequent exceptional measurement opportunities. Here, we emphasize CMB-S4's impact on  two key themes of particular relevance to the science of cosmic acceleration (for more details see discussions in white papers~\cite{CMB-S4:2022ght,Chang:2022tzj,Baxter:2022enq,Achucarro:2022qrl,Amin:2022soj,Blazek:2022uzw}). 
\begin{itemize}
\item \textbf{Primordial gravitational waves and inflation}:  Cosmic inflation is a prominent theory for the origin of structure in the Universe. A detection of primordial gravitational waves from inflation would be historic,  providing evidence for the quantization of gravity and opening a window into the very early Universe \cite{Achucarro:2022qrl}.
The factor of five leap in sensitivity and exquisite systematics control embedded in the CMB-S4 design will enable the experiment to cross major theoretically motivated thresholds through either a detection of these primordial gravitational waves from an inflationary epoch or an upper limit that will rule out entire classes of the most compelling inflationary models.  In either outcome, CMB-S4 will dramatically advance our understanding of the primordial Universe. 

\item \textbf{The dark Universe}:  CMB-S4 will also provide multiple compelling probes of the late-time universe which will enable stringent tests of dark energy and other models of the Universe's observed accelerated late-time expansion.  These probes include precision measurements of the gravitational lensing of the CMB, the kSZ velocity field, and a large ($>$100,000) sample of massive galaxy clusters discovered via the tSZ effect. There is an additional wealth of information to be gained through cross survey analyses between the CMB and other tracers of structure as detailed below.  

\end{itemize}

In addition to the fundamental physics above, the sensitivity and sky coverage of the CMB-S4 millimeter-wavelength survey will enable other important scientific opportunities in the themes of `mapping matter in the cosmos' and the `time-variable millimeter-wave sky'.  Light relics are a well-motivated potential contributor of energy density in the Universe that lead to an observable signal in the CMB temperature and polarization~\cite{Dvorkin:2022jyg}.   CMB-S4 will be able to constrain the effective number of neutrino species with a sensitivity to Weyl fermion and vector particles that froze out in the first fractions of a nanosecond.
For explorations of the cosmological and astrophysical science of the growth of structure, maps of the ionized gas distribution at CMB-S4 sensitivity will lead to the detection of an order of magnitude more high-redshift ($z > 2$) galaxy clusters than found by Stage III experiments~\cite{cmbs4:2019}.  This is just one example of the scientific potential of the ionized gas map; several others, including opportunities for complementarity, are described in the CMB-S4 white paper and its references~\cite{CMB-S4:2022ght}.   Finally, CMB-S4 will provide new, key insights into millimeter-wavelength transient phenomena by making a repeated, systematic survey of a larger area of the sky at a cadence of approximately a day.  Limited studies of the variable millimeter-wave sky exist, and therefore the  CMB-S4 survey will open this discovery space.

\subsection{CMB-HD}

CMB-HD is a proposed CMB experiment that would have three times the total number of detectors as CMB-S4 and about six times the resolution of current and planned high-resolution CMB telescopes, opening a new regime for millimeter-wave science~\cite{CMB-HD:2022bsz}. CMB-HD would cross important thresholds for improving our understanding of fundamental physics, including the nature of dark matter and dark energy, the light particle content of the Universe, the mechanism of inflation, and whether the early Universe has new physics beyond the Standard Model, as suggested by recent H$_0$ measurements. The combination of CMB-HD with contemporary ground and space-based experiments would also provide countless powerful synergies.

The concept for the CMB-HD instrument is two new 30-meter-class off-axis crossed Dragone telescopes located on Cerro Toco in the Atacama Desert~\cite{CMB-HD:2022bsz, Sehgal:2019ewc, Sehgal:2020yja}. Each telescope would host 800,000 detectors (200,000 pixels), for a total of 1.6 million detectors. The CMB-HD survey would cover half the sky over 7.5 years. This would result in an ultra-deep, ultra-high-resolution millimeter-wave survey over half the sky with 0.5~$\mu$K-arcmin instrument noise in temperature (0.7~$\mu$K-arcmin in polarization) in combined 90 and 150 GHz channels and 15-arcsecond resolution at 150 GHz. CMB-HD would also observe at seven different frequencies between 30 and 350 GHz for mitigation of foreground contamination. 

CMB-HD would be able to measure the dark energy equation of state with an uncertainty of $\sigma(w_{0})= 0.005$ by combining galaxy cluster abundance measurements, galaxy cluster lensing measurements, and measurements of the primary CMB power spectra~\citep{Raghunathan:2021zfi, Raghunathan:2021tdc}.  This would provide a constraint on the dark energy equation of state to sub-percent level accuracy.  CMB-HD would also constrain an epoch of inflation in several ways. CMB-HD would probe the existence of inflationary magnetic fields in the early Universe via tight constraints on anisotropic birefringence. It would have the sensitivity to obtain a $1\sigma$ uncertainty on the strength of scale-invariant inflationary magnetic fields, $B_{\rm SI}$, of $\sigma(B_{\rm SI})=0.036~\mathrm{nG}$, which is below the $0.1\,\mathrm{nG}$ threshold required for inflationary magnetic fields to explain the $\mu\mathrm{G}$ level magnetic fields observed in galaxies today~\cite{Mandal:2022tqu}. CMB-HD will therefore have the capability to detect inflationary magnetic fields with about $3\sigma$ significance or greater, and such a detection would provide compelling evidence for inflation.

The cross correlation of CMB-HD with galaxy surveys would also provide powerful constraints on inflation.  CMB-HD would measure primordial local non-Gaussian fluctuations in the CMB, characterized by the parameter $f_\mathrm{NL}^{\rm local}$, with an uncertainty of $\sigma(f_\mathrm{NL}^{\rm local}) = 0.26$, by combining the kinetic Sunyaev-Zel'dovich (kSZ) signal from CMB-HD with an over-apping galaxy survey such as from the Vera Rubin Observatory. This constraint is limited by the galaxy sample from Rubin Observatory, rather than by CMB-HD, and a combination with future even higher resolution galaxy surveys would lead to even better constraints.  Reaching a target of $\sigma(f_\mathrm{NL}^{\rm local}) < 1$ would rule out a wide class of multi-field inflation models, shedding light on how inflation happened~\cite{Alvarez:2014vva,Smith2018,Munchmeyer:2018eey,Deutsch2018,Contreras2019,Cayuso2018}.  Moreover, the combination of the kSZ effect from CMB-HD with the Rubin Observatory galaxy survey can constrain the primordial trispectrum amplitude, $\tau_\mathrm{NL}^{\rm local}$, with $\sigma(\tau_\mathrm{NL}^{\rm local})<1$~\cite{AnilKumar:2022flx}.  CMB-HD also can provide an independent constraint on primordial gravitational waves with an uncertainty of $\sigma(r)=0.005$ via the combination of the polarized Sunyaev-Zel'dovich effect from CMB-HD with Rubin Observatory galaxies~\cite{CMB-HD:2022bsz}.  For further details see~\cite{CMB-HD:2022bsz} and~\href{https://cmb-hd.org}{https://cmb-hd.org}.

\section{Opportunities from Cross-survey Analyses}
\label{sec:cross}

The next decade will see dramatic improvements in our ability to probe the Universe, with major leaps in capabilities occurring nearly simultaneously across many new facilities.  Each of these new facilities will enable transformative science, but joint analyses of the resultant datasets will be more powerful and robust than what can be achieved with any individual instrument.  Notably, cross-survey analyses will improve the constraints on cosmic acceleration that drive the design and requirements for cosmological surveys into which DOE has invested, and also leverage those investments to constrain other aspects of fundamental physics that are  important for our understanding of the Universe.  At present, however, cross-survey analyses can be challenging to initiate, organize and fund.  We therefore advocate for the creation of clear pathways to support cross-survey analyses as part of the core mission of the DOE Cosmic Frontier. 

\subsection{Static Probes}

\begin{figure*}[t]
    \centering
    \includegraphics[scale=0.65]{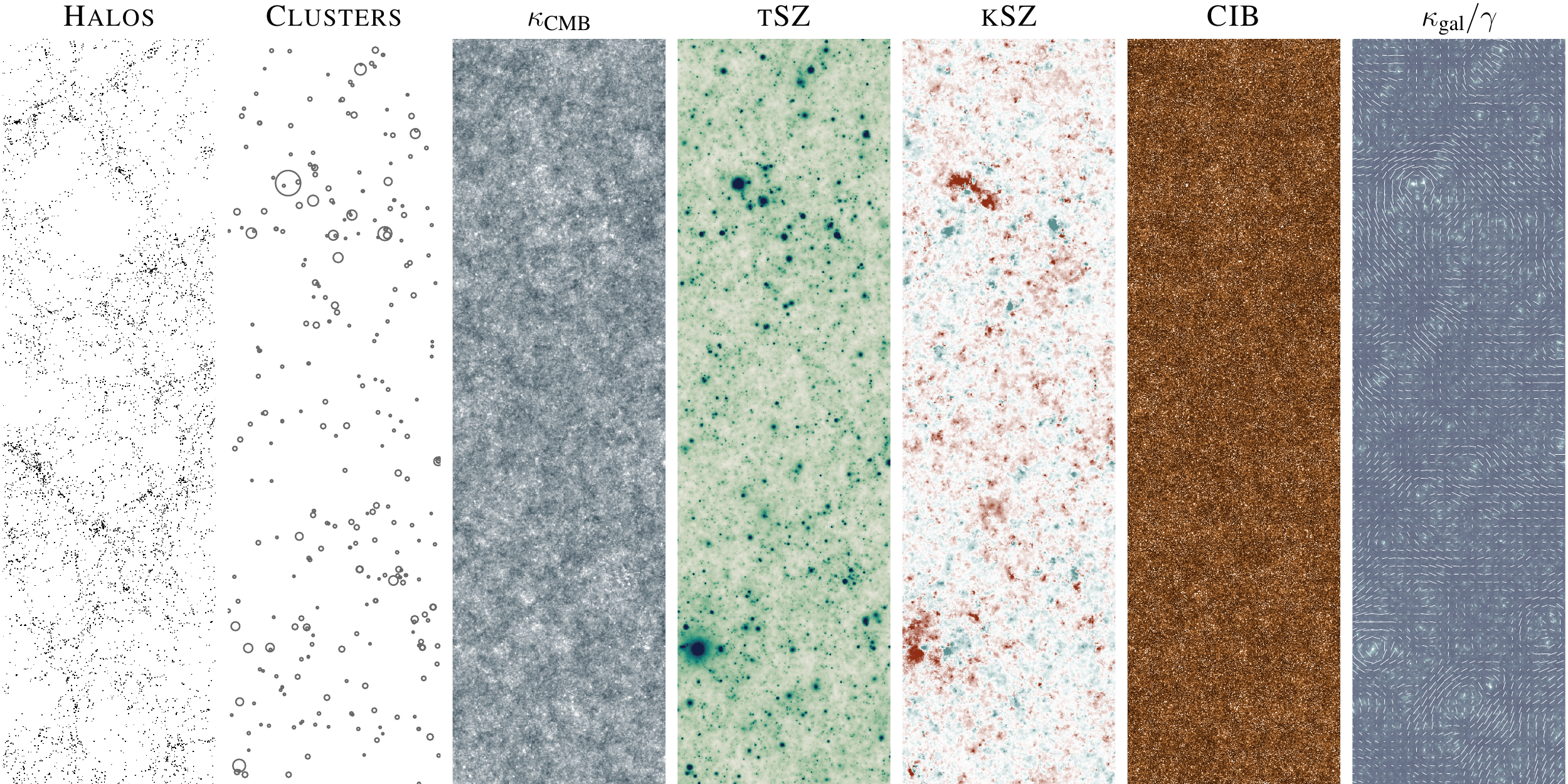}
    \caption{Simulated maps of the same patch of the Universe, as measured with several different cosmological probes (from left to right): dark matter halos (detectable via the galaxies they host), galaxy clusters (with the size of the circles indicating the cluster mass), gravitational lensing of the CMB ($\kappa_{\rm CMB}$), the thermal Sunyaev Zel'dovich effect (tSZ), the kinematic Sunyaev Zel'dovich effect (kSZ), the cosmic infrared background (CIB), and gravitational lensing of galaxy shapes (shading indicates the convergence, $\kappa_{\rm gal}$, while white lines indicate the shear, $\gamma$).  Although each probe is very different, they are all sourced by the same underlying large-scale structure, and are therefore correlated.  Joint analyses of these different probes can yield access to new cosmological information about the underlying structure. Simulated data from Omori (in prep.). }
    \label{fig:multiwavelength}
\end{figure*}

We first consider cross-survey analyses between ``static'' probes of the Universe, i.e. those observables that do not change significantly over the time frame of a survey.  This includes probes like galaxy positions, weak gravitational lensing, and the Sunyaev Zel'dovich effect.  Current and future cosmic surveys will obtain measurements of multiple static probes that overlap over significant fractions of the sky.  Such measurements will enable many cross-survey analyses to obtain tighter and more robust constraints on the fundamental ingredients of our Universe.  We illustrate the diversity and complementarity of overlapping cosmic probes in Fig.~\ref{fig:multiwavelength}.

By combining overlapping probes from different surveys, new information about cosmological structure can be extracted, and the cosmological constraints from individual surveys can be made more robust to possible systematic biases. Some prominent examples include: 
\begin{itemize}
    \item {\bf Improved cosmological constraints.} By leveraging multi-wavelength data, combining imaging and spectroscopic surveys, cross-survey analyses will improve cosmological constraints from the evolution of large-scale structure.   
    \item {\bf Improved robustness of cosmological constraints}. Analyses of cross-survey correlations help to isolate survey-specific systematic effects and break degeneracies between cosmological parameters and nuisance parameters, making cosmological constraints more robust. In addition, multi-wavelength data allow for improved understanding of baryonic processes, 
    one of the main sources of systematic uncertainty in cosmological analyses of large-scale structure. 
\end{itemize}

Measuring cross-correlations between different cosmological probes requires overlapping measurements on the sky.  The survey strategies of several operational and planned DOE-funded cosmic surveys  --- including optical imaging, spectroscopic, and CMB surveys --- have significant overlap.  The potential therefore exists to harness the power of cross-correlations between them. However, modeling multi-survey correlations necessarily requires additional work beyond that typically undertaken by single surveys. In particular, there are significant technical challenges in simultaneously modeling and simulating observables that span a wide range of wavelength and scales, and that involve multiple astrophysical processes.

Beyond the technical challenges associated with cross-survey analyses, there are also practical difficulties associated with this work.  Any such analysis necessarily requires detailed knowledge of data products generated by multiple surveys.  Some of this information may be proprietary, and not easily shared.  Previous cross-survey analyses have typically waited until data products become public (thereby delaying results) or have operated through cross-survey memoranda of understanding (MoU).  Relative to single-survey analyses, analyses conducted through MoU are often subject to additional bureaucratic hurdles that can delay progress and unnecessarily increase workloads.  These difficulties can be significant enough to discourage cross-survey analyses, a clearly suboptimal outcome.  

To capitalize upon these opportunities and address the associated challenges, a qualitatively new level of investment in cross-survey, joint-probe infrastructure is required -- this includes simulations, associated modeling, coordination of data sharing, survey strategy, and training for the next generation of scientists in a way that transcends any individual project or collaboration. The required investments are substantial, but they are critical for the next generation of cosmic surveys to fully realize their potential. Below we present a summary of future opportunities for growth that have potential to multiplicatively enhance the scientific returns of cosmological surveys in the 2020s:

\begin{itemize}
    \item \textbf {Joint simulations:} Nearly all of the multi-probe analyses discussed above require high-fidelity synthetic data that is validated against observational data. The computational demands of these simulations can be high, and an intensive human effort is required in order to generate synthetic data that is of sufficiently high quality to merit this expense. Considerable progress has been made in this area in recent years, but efforts are typically limited to an individual survey, or even an individual probe in isolation. For example, most CMB simulations do not include physically realistic models of galaxy populations at low redshift, and synthetic datasets tailored for optical surveys of galaxies do not commonly include realistic treatments of the diffuse gas that can be observed in CMB surveys via, e.g., the SZ effect.
    As a result, the need is increasing for simulations that are suitable for multi-wavelength cross-correlation analyses. Addressing this widespread need is a key opportunity for further growth in the area of generating multi-survey synthetic data, and the wider cosmology community stands to greatly benefit from increased support for these efforts.
    \item \textbf{Joint modeling and analysis:} Current toolkits such as \texttt{Cobaya} \cite{Cobaya}, \texttt{Monte Python} \cite{MontePython}, \texttt{CosmoLike} \cite{cosmolike}, and \texttt{CosmoSIS} \cite{Zuntz:2015} have been successful in combining a number of ``standard'' large-scale structure probes in Bayesian analyses. Sophisticated modeling efforts with capability to make multi-wavelength predictions are commonly implemented in custom codebases that require highly specialized techniques in order to infer cosmological parameters in a Bayesian fashion. Fully integrating a new generation of models together with cosmological inference pipelines is another exciting opportunity, and would leverage new technologies such as machine learning methods, GPU interfaces, automatic gradient approaches, and likelihood-free inference methods. 
    \item \textbf{New initiatives enabling joint analyses:} By construction, multi-survey analyses in the era of large collaborations are not hosted under one single collaboration with well-established communication structure and analysis tools. Presently such analyses are enabled by MoUs and other agreements, or carried out with public data. This structure can create an inherent barrier for multi-survey analyses, and suppress potential opportunities for exciting discoveries.  Conversely, new levels of effort in cross-survey collaboration could offer major benefits to the scientific returns of future surveys. Such initiatives could include coordination of survey strategy to ensure overlap, joint processing of data, and coordination of cross-survey blinding strategies. New funding lines that focus on multi-survey cross-correlation analyses could be an effective, modest way to address some of these limitations.  The scope of these problems, however, warrants consideration of new ``centers'' focusing on development of joint simulation/modeling/analysis tools, as well as training/education for the next generation of cosmologists who will be confronted with data already in the 2020s that is of a qualitatively new character from previous decades. 
    In addition, this effort should be combined with a support for a healthy and equitable collaboration community \cite{2022arXiv220601849A,2022arXiv220403713H}.
    \item \textbf{Support for proposed cosmic survey instruments:} The enormous potential of joint analyses discussed in this white paper is necessarily built on the success of single-probe experiments.  Enabling cross-survey analyses requires support for wide-field cosmic surveys including those listed in Figure~\ref{fig:roadmap}, and many more described in accompanying Snowmass white papers \cite{CMB-HD:2022bsz,2022arXiv220307291D,2022arXiv220306200C, 2022arXiv220307258K, Ferraro:2022, Schlegel:2019eqc}.   In return, joint-probe analyses will provide critical and complementary information for understanding cosmic acceleration and other fundamental physics.
\end{itemize}

\subsection{Transient Probes} 

Transient science is a key frontier of modern cosmology, with profound implications for our understanding of dark energy, cosmological distances in the Universe, extreme strong-gravity environments, and high-energy physics. An extensive variety of transient science requires diverse data sets that can only be acquired via multiple experiments and surveys. For example, optical telescopes are necessary for the search and association of transient counterparts of gravitational-wave standard sirens detected by gravitational-wave observatories to measure the Hubble constant $H_0$ \cite{2009PhRvD..80j4009C,AbbottH0,2019ApJ...876L...7S,2020ApJ...900L..33P,2021arXiv211103604T,palmese_StS_DESI,Mukherjee:2022afz}. Moreover, studies using transients in combination with data from neutrino experiments such as IceCube have been proposed for measurement of the neutrino masses \cite{Pagliaroli:2011zz,Nakamura:2016kkl}. 

To measure the properties of dark energy specifically precise and accurate distance measurements will be needed for the Rubin Observatory Type Ia supernovae via spectroscopic, near-infrared, and enhanced temporal sampling \cite{snpv}. 
A high-efficiency search and discovery program will also be needed for the electromagnetic counterparts of standard sirens, to enable a measurement of the Hubble constant that is independent from the systematic uncertainties affecting other dark energy probes~\cite{Kim:2022iud}. One can also test theories of gravity from GW sources for both bright and dark standard sirens~\cite{Mukherjee:2020mha}. High spatial resolution and enhanced temporal sampling are also required to obtain precise time delays by modeling strongly lensed systems discovered by Rubin Observatory, and therefore, independently measure the Hubble constant~\cite{Linder:2011,TreuMarshall2016,Wong:2020,BirrerTreu:2021}. Finally, peculiar velocities inferred from the distances of standard sirens and supernovae could be compared with the density perturbations within the DESI survey volume to measure the strength and length scale of gravity~\cite{palmese20,Diaz:2021pem,Mukherjee:2020hyn}.

Currently a critical issue experienced by the HEP community is the perceived inconsistencies between different experiments and/or cosmological probes. A prime example is the Hubble tension, where the Hubble constant measured from the cosmic microwave background, baryon acoustic oscillations, and Type Ia supernovae are not in agreement. These tensions present an opportunity for our community to make a breakthrough in our understanding of dark energy.   Their resolution may lie in new fundamental physics, or unaccounted-for systematic errors. Transient science can play a crucial role in solving this challenging issue with enough resources and support for developing its full potential (see Section 2 of~\cite{Kim:2022iud}, for example).

\textit{No experiment alone can solve the dark energy problem. Such a breakthrough will require a complex network of experiments, small and large, working in tandem. As dark energy is a priority of our community, it is natural that we ramp up our efforts to build and operate those experiments, optimizing for dark energy science. Those efforts include near-, medium-, and long-term investments. For example, we need data from gravitational wave observatories, and from telescopes that can identify their transient counterparts and host galaxies. Therefore, supporting partnerships between ongoing projects (such as DES/DESI/LSST and the LIGO/Virgo/KAGRA Collaborations) as well as the development of a third-generation gravitational wave observatory (e.g. Cosmic Explorer~\cite{2022arXiv220308228B}), which until recently had been considered as a outside the scope of the HEP community, is consistent with our goals.}

Time-domain science with multiple experiments have unique considerations that do not occur for self-contained experiments, e.g., regarding experimental design. In a multi-experiment context, experimental designs can be optimized for a joint rather than stand-alone project. The joint analysis of low-level data products (e.g., pixels) can preserve significantly more information than the combination of lossy final data products. To benefit from this kind of joint analysis, static and time-domain resources are necessary for developing a new infrastructure for real-time communication between experiments.

New support is needed to enable this time-domain science to achieve and surpass the precision level of the current standard static experiments. As such, analysis of multiple experiments requires resources beyond the sum allocated to the individual ones. We need to develop simulations that account for different probes to support self-consistent interpretation of the multi-experiment data. Ultimately, new experiments must be developed and supported when existing ones are insufficient.

We advocate for these transient science initiatives detailed in Kim et al.\ (2022)~\citep{Kim:2022iud}:
\begin{itemize}
    \item Small projects to acquire supplemental data to enhance the science reach of transients discovered by Rubin LSST.
    \item Use of the 4-m Blanco telescope hosting DECam  for fast and effective search and discovery of transients including gravitational wave events, strongly lensed quasars and strongly lensed supernovae.
    \item Infrastructure that enables cross-experiment, cross-facility coordination and data transfer for time-domain astronomical sources.
    \item Theory/modeling that improves understanding of the transient astrophysical probes that are used to study cosmology.
    \item A US-HEP multi-messenger program, supported with dedicated target-of-opportunity allocations on US-HEP and partner facilities for the follow-up of gravitational waves and rare neutrino events.
    \item The development of a novel standard siren survey program using next-generation gravitational wave observatories to fully incorporate this new observable into the research portfolio for dark energy science. 
     \item Construction of novel large-scale projects for a multi-messenger dark energy survey, including gravitational wave observatories and optical NIR telescopes, designed to resolve the current tensions and advance understanding of dark energy and cosmic acceleration.

\end{itemize}
\section{Small Projects and Pathfinders}
\label{sec:smallproj}

In 2016 and 2017, the community held two workshops to discuss future opportunities for survey science and to develop a small-project portfolio that would include technology developments to enable a major new Stage V Spectroscopic Facility. The findings are summarized in Ref.~\cite{Dawson:2018fob}. In the following, we provide an overview of the findings that are relevant in particular to the development of new facilities to explore cosmic acceleration.

\subsection{Spectroscopy Pathfinder}

In Ref.~\cite{Dawson:2018fob} the importance of new technology developments were highlighted. These developments are needed in the near future to enable a credible design for a Stage V spectroscopic facility. In particular, near-term investigations of the following areas will be crucial:

\begin{itemize}
    \item {\bf Detector technologies} to extend to higher redshift (e.g., Germanium CCDs) and lower noise (e.g., Skipper CCDs). Current silicon CCD detectors have a wavelength cutoff due to the band gap of silicon. Lower band gap materials, such as Germanium offer the potential to extend to higher redshift.     Precision measurements of faint, distant sources can be dominated by detector readout noise. Novel Skipper CCD detectors offer the ability to reduce noise through multiple non-destructive measurements of the charge in each detector pixel. A challenge in Skipper CCD technology is the readout time, which scales with the number of non-destructive measurements that are made.
    \item {\bf Fiber positioner technologies} to enable smaller pitch, denser packing, and greater robustness. Two technologies are currently considered for fiber positioners. The robotic twirling post design has been used by DESI. R\&D is ongoing to shrink the patrol radius and increase the packing density. Robustness is a current challenge faced by twirling post technology. The second technology is tilting spines, which are being used by the 4MOST  spectrograph. R\&D is ongoing to shrink the pitch and demonstrate precise control of fiber positions. 
    \item {\bf Wide-field optics} to enable larger focal planes that can hold more fibers. This is a critical component toward increasing total fiber number. Advances have been made in the context of several telescope designs to allow $> 1$-meter diameter focal planes (i.e., MegaMapper, MSE, SpecTel). Current challenges are the fabrication of large-diameter lenses.
    \item {\bf Verification of high-redshift target viability} (e.g., Lyman-alpha emitters, Lyman-break galaxies, etc.). This work is currently on-going with targeted observations by DESI.
    \item {\bf Narrow-band targeting} would use large-field imagers outfitted with multiple medium- or narrow-band filters to improve targeting efficiency for future spectroscopy. Such a campaign could be executed by DECam outfitted with a new set of filters for a moderate cost.
\end{itemize}

\subsection{21-cm Pathfinders}
Neutral hydrogen is ubiquitous in the Universe after the CMB was formed, such that its 21\,cm emission can trace large-scale structure across cosmic time. At low redshift, maps of the 21\,cm emission line can form a galaxy survey to constrain models of dark energy. At higher redshifts, they can improve measurements of the primordial power spectrum as a probe of inflation (described in the CF5 report). In all cases, the primary challenge is removing bright foreground emission from the resulting maps, which drives the instrument design.

Maps of 21\,cm emission at redshifts $z<6$ form a galaxy survey using the signal from neutral hydrogen trapped in galaxies. Unlike their optical counterparts, these radio surveys naturally have wide fields of view and observe all redshifts in their band simultaneously, allowing these radio telescopes to quickly survey very large volumes spanning the redshift desert ($z\sim 1$--$3$) and beyond ($z\sim 3$--$6$), where optical spectroscopy is challenging or impossible. To detect cosmological neutral hydrogen across a wide redshift range and target inflation and dark energy science goals, a dedicated 21\,cm instrument will require a close-packed array of thousands of dishes at least 6\,m in diameter across a wide redshift range~\cite{2018arXiv181009572C, PUMAAPC}, resulting in a radio array with a physically large footprint ($\sim$ km scales) that requires efficient signal transfer and an extremely large digital correlator.  

Dedicated experiments to use 21\,cm emission to map structure have shown that the primary challenge is foreground removal~\cite{CHIMEresults,2015ApJ...809...62P,2019ApJ...883..133K,2019ApJ...887..141L,2016ApJ...833..102B,2016MNRAS.460.4320E}, which drives requirements for instrumentation calibration and design. Solving these design challenges requires targeted R\&D for a pathfinder that has uniform elements; a well-controlled bandpass; instrument stability and stabilization methods using digital signal processing and fast real-time analysis; robust real-time RFI flagging; new calibration techniques for beam and gain measurements potentially including drone-based calibration; and requires analysis and simulations to fold in calibration measurements and assess their impact on cosmological parameter estimation\cite{2018arXiv181009572C}. The primary US pathfinder targeting this R\&D is The Packed Ultra-wideband Mapping Array (PUMA)~\cite{PUMAAPC}, a proposed next-generation 21\,cm \ intensity mapping array which is optimized for cosmology in the post-reionization era. The reference design calls for PUMA to consist of a hexagonal close-packed array of 32,000 parabolic dishes 6m in diameter, observing at 200-1100\,MHz, corresponding to a redshift range of $0.3 < z < 6$. The pathfinder array for this experiment is the PUMA-5K array, a staged deployment of 5000 dishes that would be used to test the analog, digital, and calibration equipment at a scale large enough to assess success on the sky. Specific technology R\&D required includes:
\begin{itemize}[nosep]
    \item Digital electronics at or near the dish foci.
    \item A timing distribution network that spans kilometers with relative timing accuracy better than a picosecond.
    \item Real-time data processing, including real-time calibration, to enable  essentially real-time data compression across interferometer inputs.
    \item Analog system design that includes uniformity of all elements and smooth response across a wide bandwidth.
\end{itemize}

Finally, the Dark Ages ($150 < z < 20$) are a particularly clean probe of the primordial power spectrum and its statistics, including searches for non-Gaussianity. However, measurements during this era are extremely challenging because the resulting long wavelengths ($\lambda \simeq $7 to 70\,m) require a physically large instrument and must contend with non-negligible effects from the Earth's ionosphere and significant contamination from human-generated radio sources (RFI). To assess whether the far side of the moon is adequate to address these issues, the DOE and NASA are collaborating to launch the pathfinder experiment LuSEE-Night (Lunar Surface Electromagnetics Experiment at Night) in lateW 2025 to deploy 4 steerable monopole antennas to characterize the radio sky at frequencies 1-50MHz with percent level absolute calibration and a $10^{-3}$ relative calibration between frequency bands. With data collected over 12 nights, it should provide measurements of the low-frequency radio sky below 50~MHz, demonstrate the feasibility of Dark Ages cosmology from the far side of the Moon, should have sufficient sensitivity to exclude presence of a monopole signal at about the 1 Kelvin level, about 1-2 orders of magnitude above the expected signal yet sufficient to constrain some models predicting non-standard properties of baryon thermodynamics during the Dark Ages.

\subsection{Line-Intensity Mapping}
Line-intensity mapping (LIM) is a nascent technique for mapping the large-scale structure (LSS) in the universe by measuring the spatial distribution of an atomic or molecular emission line with low-resolution spectrometers ($\lambda / \Delta \lambda < 1000$)~\cite{Kovetz:2017agg,Karkare:2022bai}.
The ability to measure multiple emission lines over a wide range of redshifts $z>2$, beyond the range of current galaxy surveys, makes LIM a particularly promising technique for future surveys of large-scale structure.
Although this method can be used with any emission line, LIM using mm-wavelength tracers, such as the rotational transitions of CO and the [CII] ionized carbon fine structure line, is of great experimental interest because such emission can be detected over the redshift range of $0<z<10$ from the ground using technology that is already in widespread use in CMB and sub-mm telescopes.
In addition, the Galactic foregrounds are significantly less bright in these frequency ranges than in 21cm surveys using similar techniques.

LIM with mm-wave tracers may be capable of making very significant improvements in constraints on primordial non-gaussianity, neutrino properties, light thermal relics, and dark energy, but doing so will require experiments with significantly more receiver elements and longer integration times than  currently exist and development of sophisticated analysis pipelines.
A suite of small projects, including CCAT-p, COMAP, CONCERTO, EXCLAIM, mmIME, SPT-SLIM, TIM, and TIME, is currently prototyping various spectrometer and detector technologies, at the scale of $10^5$ spectrometer-hours or less.
By contrast, constraining the amplitude of local-type primordial non-gaussianity to a level $\sigma(f_\textrm{NL}) \sim 1$ that would distinguish between single- and multi-field inflation, or dtecting the minimal sum of the neutrino masses at $5\sigma$ would require a survey with $10^8$ spectrometer-hours, three-orders of magnitude larger than existing projects.
To reach this level of sensitivity requires investment in a program of technology development, complemented by the staged deployment of projects with increasingly large focal planes of detectors to demonstrate these technologies in the field, analogous to the way the CMB field has grown from few-pixel experiments to an experiment like CMB-S4 with 500,000 detectors.  Concurrent, steady improvement in modeling, analysis techniques, tools and pipelines is a must.
Specific technological capabilities to develop include:
\begin{itemize}[nosep]
    \item \textrm{{\bf On-chip spectrometers:}}
    A key challenge in scaling mm-wave spectrometers to very high channel counts is the spectrometer element itself.
    Traditional technologies, such as diffraction gratings, Fourier Transform or Fabry-Perot spectrometers, and heterodyne detection perform well for the existing generation of small focal planes, but each has difficulties scaling to larger focal planes.
    On-chip spectrometers, which channelize the incident radiation using a filter bank on the same silicon wafer as the pixel itself (similar to the current generation of multichroic CMB detectors, but with many more channels), offer a promising solution to the scaling problem by shrinking the physical size of the spectrometer and eliminating complex coupling optics between the telescope and the pixel.
    Despite these attractive features for mm-wave LIM, on-chip spectrometers are comparatively less mature than traditional technologies, and require field demonstration to test existing architectures and adapt the form factor to more efficiently use focal plane area of telescopes.
    \item \textrm{{\bf Multiplexed readout electronics:}}
    Spectrometers with $10^2 - 10^3$ spectral channels per spatial pixel require far more detectors or channels than broadband cameras.
    Increased multiplexing factors are essential in order to reduce the per channel cost of the readout system to a manageable level.
    Advances in FPGA technologies, such as RF system-on-a-chip (RFSoC) devices, for example, may reduce per channel cost of readout for kinetic inductance detectors to the level of \$1--2 / channel.
    \item \textrm{{\bf Telescopes and facilities:}}
    Detectors for mm-wave LIM, especially on-chip spectrometers, are compatible with the existing generation of small- and large-aperture telescopes built for CMB observations, including SPT, ACT, SO, and CCATp.
    In some cases, these existing facilities can be used to host mm-wave demonstration cameras without compromising other science goals (e.g. SPT-SLIM on SPT and PrimeCam on CCATp).
    A staged deployment of mm-wave LIM cameras of increasing size, using existing telescope infrastructure, is critical for achieving on-sky demonstrations of detector and readout technologies and prototyping analysis pipelines.
\end{itemize}

Since mm-wave LIM is still a very young field, a staged program of surveys of increasing size will provide valuable data sets for developing analysis techniques and characterizing observational systematics.
For example, the problem of interlopers --- lines from different transitions and redshifts that map to the same observed frequency ---is one well-known systematic with several proposed solutions, but these mitigations have yet to be tested on real data.
Similarly, the effect of atmospheric lines at mm-wavelengths is not expected to corrupt cosmological LIM signals, but projecting to the low required noise levels is difficult.

\section{Multi-Messenger Probes}
\label{sec:multimessenger}

\subsection{Gravitational Wave Observatories}

Historically, gravitational wave (GW) observatories were outside the scientific scope of the US HEP community’s efforts. However, since the discovery of GW150914 by the LIGO \& Virgo collaborations and the realization that gravitational wave standard sirens are a powerful dark energy probe (e.g., GW170817), the community has embraced this type of experiment. For that reason, we incorporate them here in the discussion of this report. The  next decade will see upgrades of existing facilities, as well as developments of new large-scale projects. Both are discussed below.

\subsubsection{Current Ground-Based GW Facilities}

Currently, there are two LIGO facilities, in Livingston, Louisiana (LLO) and Hanford, Washington (LHO). Each of these detectors has 4-km long arms and is expected to have sensitivity for binary neutron star (BNS) mergers out to 160--190\,Mpc during the LIGO Fourth Observing run (O4). Other facilities expected to be online during O4 and beyond are Virgo, in Italy, as well as the recently constructed KAGRA in Japan. Each of these facilities has 3km long arms and is in various stages of sensitivity. During O3, Virgo reached a BNS range of 40--50\,Mpc and is expected to ramp up to 80--115\,Mpc during O4. KAGRA, on the other hand, will be online only  for a portion of O4 and it is expected to reach $\sim 10$\,Mpc BNS sensitivity. During the O4 run the LIGO/Virgo Collaboration (LVC) expects to detect $10^{+52}_{-10}$ BNS events.

\subsubsection{Future Ground-Based GW Facilities}

With the numerous GW discoveries in recent years, plans for new ground-based facilities are already underway. LIGO-India has been approved for construction and should be operational by the end of the decade. This detector is planned to be the same size and design as the current LIGO facilities and will come online at similar sensitivity as current detectors~\cite{2019arXiv190402718C}. The addition of LIGO-India will greatly improve the localization of GW events, as well as help to measure the polarization of GWs. Additionally, plans for the Einstein Telescope have been moving forward~\cite{ETCern}. This facility would be underground with 10-km long arms and would be a third-generation (3G) GW observatory. In the US, Cosmic Explorer (CE) is the current 3G proposal for the 2030s, and it is now in the conceptual design phase~\cite{2021arXiv210909882E}. One of the CE’s proposals is two detectors of 40-km long arms that will be able to reach sources at $z \sim 100$ in network with the Einstein Telescope. 

\subsubsection{Future Space Based-GW Facilities}

Space-borne gravitational wave observatories are being planned or proposed for the 2030s. The Laser Interferometer Space Antenna (LISA), a constellation of three spacecraft forming an equilateral triangle with sides 2.5-million km long, is understudy. LISA is led by the European Space Agency, but with significant contributions from NASA and the US, along with several other countries. LISA will open a new window in the GW spectrum by detecting sources in the mHz frequency band. Its main detections will be the inspiral and merger of massive binary black holes (MBBHs), with masses ranging between $10^4$  and $10^7 M_{\odot}$, at redshifts out to $z \sim 10$. LISA will observe the early inspiral phase of stellar-mass binary systems months to years before they are observed in terrestrial detectors. This has the potential to open an entire new chapter of the GW field by adding the power of multi-band observations. LISA scientific objectives include measurements of the expansion rate of the universe by means of GW observations alone and further to constrain cosmological parameters through joint GW and electromagnetic (EM) observations. Another objective of LISA is to understand primordial stochastic gravitational wave backgrounds (SGWBs) and their implications for early universe and particle physics \cite{2022arXiv220405434A}. 

Complementary to LISA, the Deci-hertz Interferometer Gravitational Wave Observatory (DECIGO) is the proposed Japanese space mission in the decihertz frequency band. DECIGO consists of four clusters of three spacecrafts (LISA-like) with an arm length of 1000-km. The main goals of DECIGO are the detection of primordial gravitational waves to verify and characterize the inflationary era, measurement of the expansion rate of the universe, and to characterize dark energy, and the prediction of accurate time and direction for electromagnetic follow-up observations. DECIGO will catch $\sim 100,000$ gravitational wave events per year from neutron star binary mergers within $z \sim5$ \cite{10.1093/ptep/ptab019}. A decihertz observatory like DECIGO is projected to determine the Hubble constant to $\sim 0.1\%$, and the dark-energy parameters $w_0$ and $w_a$ to $\sim 0.01$ and $\sim 0.1$, respectively \cite{2009PhRvD..80j4009C}.





\bibliographystyle{JHEP}
\bibliography{Cosmic/CF06/myreferences}

\end{document}